# Murder by Design:

## Design thinking approach for pre-emptive cybernetic security design

Rebekah Rousi – School of Marketing and Communication, University of Vaasa, Finland

**Introduction**

Besides COVID-19, one of the ever looming threats and causes of personal and organisational stress these days is that of cyber security breaches. While increasingly it is being acknowledged that security needs to be incorporated in the design process from the ground up (Neuman, 2009) and not as an afterthought, the issue of security design is currently characterized by its cybercentricity (Ess, 1998; Gordon, 2001). That is, while paramount effort is now being placed on researching and developing systems to ensure the security of global online information, striking neglect is occurring in relation to offline systems, safety and security of the body. The problem is that while research and development (R&D) efforts are placed on securing information, a large amount of the affected areas of crime are left vulnerable. A 'Plato's Cave' effect can be witnessed in which the academic and design worlds alike, are chained to the inside of a cave, only able to understand it through the wall shadows of phenomena moving passed the mouth of the cave (digital information), not indeed through observing the phenomena itself (the world around it). Through focusing on these shadows, that is, the expanse of digital information (data) we are ignoring what else is happening within the world. Furthermore, scientists and designers are missing important information and occurrences continuously occurring within the rest of the analogue world. Human beings still live in and through an organic, social and cultural body. Motivations and intentions behind deviant thinking, the types of thinking that are deliberately dealing with online information in a corrupt way, are not purely concerned with online life alone (see e.g., Ashurst & McAlinden, 2015; Loughlin & Taylor-Butts, 2009; Maras, 2019). Rather, deviant thinkers (Internet predators) also live an embodied existence in which their needs and motivations are closely related to personal physical wellbeing and gratification (Baudrillard, 1996).

At the heart of this ideas paper lies the question: How can designers out-smart or go beyond the limits of the most creative thinkers around – deviant thinkers, who use a combination of elements within cybernetic (organic [bodily related] and artificial [online and offline]) systems to cheat and maim targets (victims)? And, how can design thinking be used as a methodological strategy to incorporate pre-emptive cybernetic security into design projects? Online and offline spaces are filled with various elements and characteristics that provide apt puzzle pieces for deviant thinkers to trap or corner, steal from, manipulate and disable targets. The term 'target' is used here due to the negative, passive and helpless image that the term 'victim' conjures (Hanish & Guerra, 2004; Papendick & Bohner, 2017). When understanding the target as a victim, scholars additionally overlook the fact that targets can be highly educated and knowledgeable in regards to matters such as security and cybersecurity and not the stereotypical 'weakest link', most often discussed. Rather, these targets are also susceptible to being entrapped through the deviant thinker's capacity for creative thinking.

**Limitations of current Cyber Security design practice and research**

Cyber security has its roots in information and communication technology (ICT) security (Von Solms & Van Nieberk, 2013). ISO/IEC 13335-1 (2004) describes ICT security as entailing all elements pertaining to the definition, achievement, maintenance of confidentiality, integrity, non-repudiation, availability, authenticity, accountability and reliability or data/information (p. 3). Conversely, data security describes the means of protecting data bound to an information system (Dhillon, 2007). Cyber security, on the other hand, is a more inclusive term that promotes a security approach that considers both the data within information systems, and the wellbeing of the humans using them – their interests, safety and privacy (Von Solms & Van Nieberk, 2013). Furthermore, it is fair to say that security in general is concerned with protecting assets in light of threats that are enabled by particular vulnerabilities (Farn et al., 2004; Gerber & Von Solms, 2005; Von Solms & Van Nieberk, 2013). Cyber security goes beyond the technology to understand the connections and vulnerabilities presented by cyber or information-based attacks within broader asset areas of human life and society – such as the people and organisations themselves and their physical property (see Figure 1).

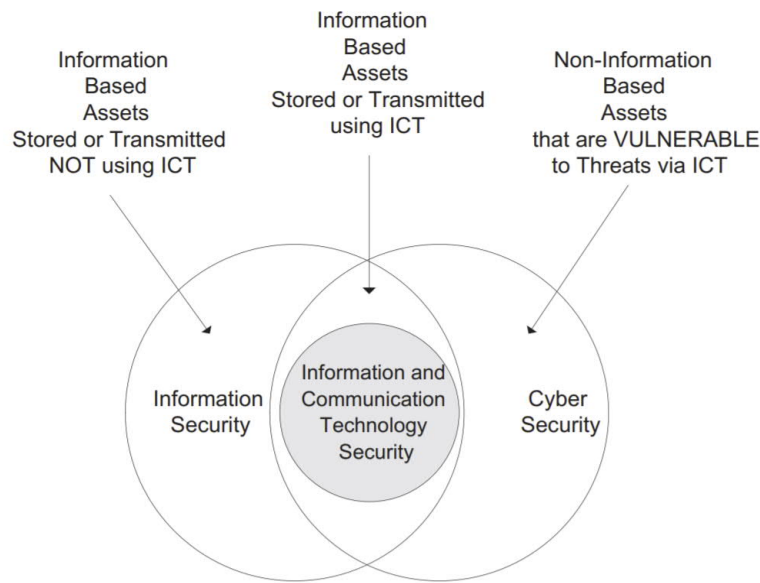

Fig. 1: The relationship between information and communication security, information security and cyber security (Von Solms & Van Nieberk, 2013, p. 101)

There are two fundamental problems however with current approaches in cyber security. Firstly, within this field researchers and designers approach security in a cybercentric, or information technology (IT) centric, manner (Ess, 1998). That is, security and related matters of IT misuse and interpersonal violation, mainly look at the crimes as being designed and undertaken via the means of information systems manipulation (Von Solms & Van Niekerk, 2013). In other words, common approaches attempt to prevent personal and organisational violation through focusing on crimes that are primarily undertaken in online spaces and/or through electronic data. This does not consider a systemic approach in which information and online spaces are merely elements, or components of a larger deviant system that aims at entangling targets in a web of interactions both online and offline. Secondly, prominent discussions and interventions aimed at examining and educating 'the weakest link' – vulnerable and careless users or stereotypical 'unknowledgeable' victims – maintain a place in the 'game' and its understanding that is at least two steps behind the perpetrator. Perpetrators are professionals and/or enthusiasts in what they do (Ferguson, 2012; Lee, 2007; Walsh, 1986). Their expert knowledge of cybernetic systems, or ways at looking at organic-artificial, online-offline elements, can be likened to that of expert chessplayers (Cowley-Cunningham, 2009; Saariluoma, 2001; Sternberg, Grigorenko & Ferrari, 2002). Criminals can see parts of the puzzle and attribute higher order meaning to networks of living and non-living elements in ways that expert chessplayers find patterns and predict consequences within every piece on a chessboard. Through focusing on the target and how they perceive their security in light of online offline systems, academics and designers are focusing on learning from the novices. In order to establish more robust approaches to securing people, their safety (physical and psychological) and their assets, more effort is needed is not simply understanding criminals and their actions, but *knowing how they think*.

**Design thinking and deviant creativity**

Design thinking and its various understandings have been popular topics in academia, design research and commercial design practice for the past few decades. As the name suggests, design thinking has literally been studied from the perspective of designer and design cognition in cognitive science and related fields. It has additionally been studied in conjunction with connected topics such as creativity, problem-solving, problem-framing and even artificial intelligence (see e.g., Carlgren, Rauth & Elmquist, 2016). Moreover, owed to multidisciplinary initiatives that have aimed to combine fields such as engineering, business, law, and design for instance, such as Stanford's d.school, design thinking is often connected to a 6-step human-centered design method that seeks to connect the designer and design process with the genuine situations and insight of stakeholders (Geissdoerfer, Bocken & Hultink, 2016; Redante et al. 2019). The method's six steps include (d.school, 2020): 1) Understand/Empathise – get to know who you are designing for by conducting interviews, learning narratives and uncovering emotions; 2) Define – reframe information and create human-centric problem statements, extract interesting revelations and tensions, add insight; 3) Ideate – brainstorm, unleash wild ideas and then start building on them; 4) Prototype – form crude objects, sketches and anticipated experiences, role play to form an understanding of the object in context with its key features, rapid materialization to learn and deliberate; 5) Test – testing the design with customers, users and other stakeholders to collect data and improve design, acquire deeper empathy and identify (embrace) failure; and 6) Assess – establishing guidelines for critically evaluating the process and product, receiving and delivering feedback then integrating feedback into the solution.

This method is customer and stakeholder focused and has on occasions been used within the context of cyber security, to empathise with stakeholders such as victims or general users (Carlgren et al., 2016). The 6-step method is often considered as been vague (Vinsel, 2018), and perhaps too prescriptive (see e.g., Jen, 2018 or even Cockton, 2020), too empathic (not enough responsibility and direction from the perspective of designer intention and design stance – see e.g., Mattelmäki, Vaajakallio & Koskinen, 2014; Crilly, 2011a; Crilly, Good, Matravers & Clarkson, 2008), and deviating from the more detailed models and theories on design thinking (Cross, 2011; Buchanan, 1992 etc.). Yet, it must be remembered that the 6-step process was developed in an attempt to combine the disciplines of law, medicine, business, mechanical engineering, humanities, the social sciences within product design. This has subsequently flowed into service and system design. The 6-step design thinking process was developed in a generalizable manner to enable the inclusion of various types of thinking (multidisciplinary thinking) into a more detailed, user-centered design structure. As a process framework, Design Thinking, can be applied across a spectrum of societal fields (Unger, 2017). Surprisingly however, it has not been studied in greater, systematic detail within the field of cyber security, or indeed, cybernetic security design. Design thinking in cyber security has been studied from the perspectives of cyber security education (Tseng et al., 2018), systems thinking (Salim, 2014) and IoT design, scenario development (Dorasmy et al., 2019) and alternative cyber security applications (Mohanty, 2019). However, to utilize design thinking from the perspective of pre-emptive cybernetic security design – with a holistic perspective of the cause-effect relations between human (organic) and internet connected technologies seems somewhat neglected.

From a general perspective, the 6-step method focuses on improving design by incorporating views of the users and other human actors within the design's ecosystem (Plattner, Meinel & Leifer, 2015; Scheer, Noweski & Meinel, 2012; Thoring & Müller, 2011). There is often emphasis on wellbeing and quality of life and interactions within the main goals and directives of the 6-step process (Chou, 2018; McDonagh & Thomas, 2010). Thus, the approach can be characterized as an 'optimistic' way of thinking through design to achieve beneficial outcomes for related stakeholders. From this perspective, design thinking has also been described by its developers as "an approach to problem-solving based on a few easy-to-grasp principles that sound obvious: 'Show Don't Tell,' 'Focus on

Human Values,' 'Bias Toward Action,' and 'Radical Collaboration' (Peter N. Miller cited in Vinsel, 2018).

Surprisingly, design thinking is still only emerging within the field of cyber security design and still has relatively little studies published on the approach. Perhaps the reason for this is due to the client-centered approach of the method that serves to place the design and development team in the position of the customer and their needs. Moreover, the prevailing 'weakest link' mentality that concentrates on the target (or victim) and their behavior seems to be continued within the cyber security design thinking approaches to date. Through overlooking the application of the 6-step model and its emphasis on empathy as a design driver, there is an opportunity missed to understand the mind, moves and creative capacity of the deviant thinker (Cropley et al., 2008; Hilton, 2010). What is proposed in this paper is the application of the 6-step model to 'empathise' with, or more aptly described, understand the thought processes, motivations and cybernetic strategies of deviant thinkers. Furthermore, while this paper emphasizes the 'understand' phase (see Figure 2), there is an aim to develop this discussion in future research to observe and test how the 6-step approach in its entirety can be utilised in specific cases to pre-emptively address and establish contingencies for scenarios that may have seen the safety and security of individuals and organisations placed at threat.

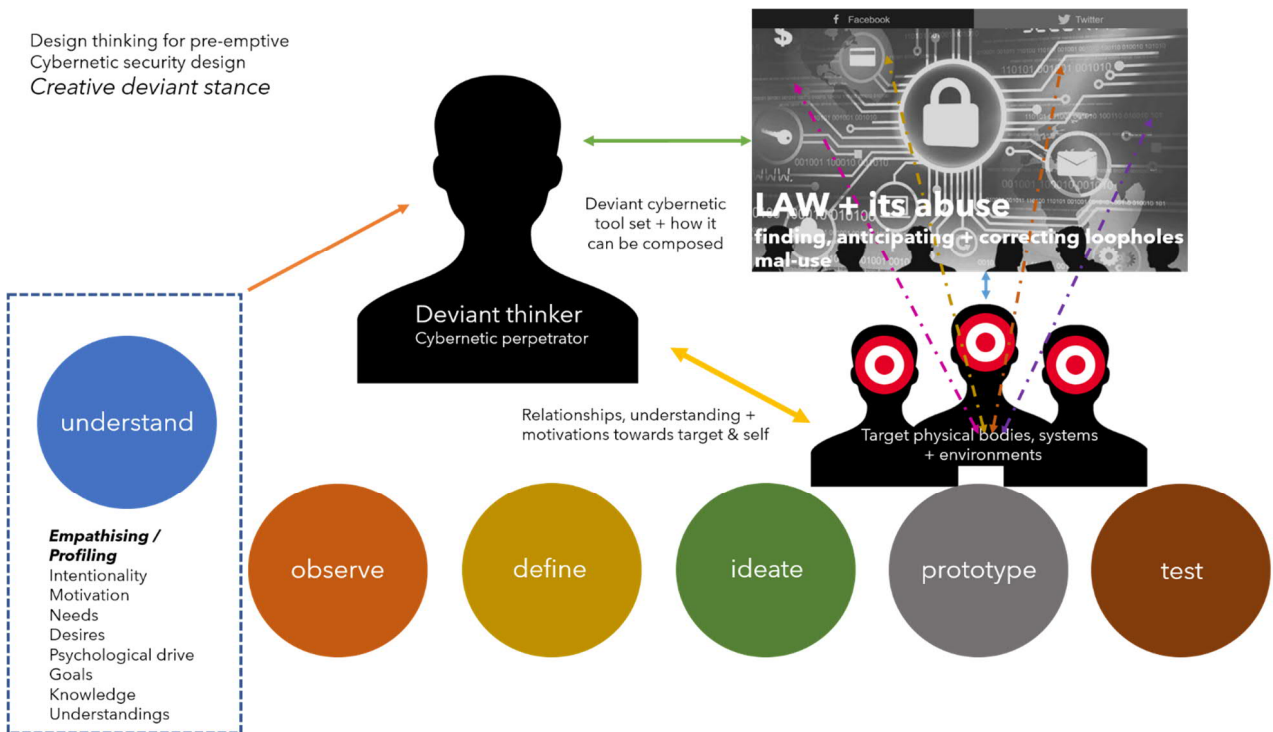

Fig. 2: Cybernetic web of deviant thinking through the 6-step model

Figure 2 provides a simplified view of the types of components designers and scholars may systematically consider and deliberate when mapping out alternative scenarios of cybernetic security threats. In order to understand the scenarios through the mind of the deviant thinker, it is important to ascertain information about: intentionality (consciousness) and intention - what someone is hoping to achieve through endeavouring on a breach in cybernetic security; motivation – what are the motivating factors for an individual or individuals to intend to undertake specific cyber security breaches; their needs (basic physio-economic needs or own psychological needs) and desires (considering gamified approaches to crime or excessive economic gain); psychological drive – predisposition to want to gain, control and manipulate others; goals – interrelated with intentions and motivations are the tangible goals of what the deviant thinker images they will gain upon successful completion of their strategies; knowledge – what knowledge is possessed and available to the deviant thinker; and understandings – how the deviant thinker understands not simply the situation, but people and property at the end of their actions, and their relationships to these. From the perspective of the design team and in relation to the cases in question, members can begin mapping out available components or systemic puzzle pieces (tools that are available for causing harm and security breach), how these may be used, and how they can directly affect the target and their safety.

**Cybernetic security and its foundations**
Not to be confused with studies in the emerging security field of cyber-physical systems (Cardenas et al., 2009; Humayed et al., 2017; Lee, 2008), that looks at the physicalisation or physical embedment of digital technology, such as seen in the Internet of Things (IoT) and embodied devices (i.e., robotics), cybernetic security considers security from the perspective of physical and virtual, analogue and digital systemic networks. One form of anticipatory (pre-emptive) cybernetic thinking can be observed in the more traditional field of insurance policy and service design. Cybernetic crime has been rife in this area through matters of insurance fraud and murder for as long as the policies have existed (Clarke, 1989; Kose, Gokturk & Kilic, 2015). Perpetrators have adopted intertextual (across texts, media and systems) modes of creative thinking that have incorporated insight of the insurance, physical factors (how they will undertake the fraud/murder) and even legal factors (once the deed is done, how can the law be used as a tool for securing the benefits?). Multi-professional teams, that can be considered 'design teams', of large insurance companies revise policies, conditions and regulations in order to minimize possibilities of insurance abuse, through for instance, value-based insurance design (Chernew, Rosen & Fendrick, 2007). There are multiple objectives in their deliberations that focus both on strengthening their competitive edge through attractive packages for their clients, while maintaining the security and stability of their own business operations and its ethical implications (Fendrick & Chernew, 2006).

There needs to be a combination of realism, apparent benefit and attainment from the perspective of the customer, in relation to *trust*ing that when they need compensation (i.e., as the result of a crime, incident or accident etc.) they will get it, yet at the same time, the understanding that there are stringent rules and regulations that may prevent the compensation from being paid out if the insurance conditions are abused in any way (Ormerod, Ball & Morley, 2012). Moreover, from the perspectives of law and ethical design, the design team needs to evaluate their decisions and packages from all of these angles, including not only how customers may benefit from policies and packages, but from the perspectives of loopholes (gaps or inadequacies within the policies) and manipulation. In order to identify the myriad of scenarios that may play out around insurance and its accompanying service design, design teams need to adopt a pre-emptive and creative approach to policy and service deliberation that anticipates human motivation, tools and devices that can be used for corrupt use of insurance, and ethical implications (Eastman, Eastman & Eastman, 1996; Tennyson, 1997). In a similar vein, cyber security, and its more holistic counterpart, cybernetic security (organic-artificial combinatory systems), has aims of design that serve to address, defend and anticipate information technology-based fraud and other security breaches (Skrynkovskyy et al., 2017).

In terms of approaching security design through the lens of cybernetics, not only the motivations and intentions of the deviant thinker may be understood, but so too are the goals. That is, a deeper understanding of desired end-goals of the perpetrator that go beyond the space of digital information, such as material outcomes, control, exploitation and violation of other people and their bodies, are included within the design process.

**Empathy, deviant thinking and intentionality – flipping sides**
Intentionality is a key composite of design, or in other words, design can be seen as the expression of intentionality and an intentional act (Crilly, 2011b; Dennett, 1989; Fodor, 1996). Intentionality is thought, mental states and mental drive towards assembling information and adjusting behaviour in a way that is aligned with underlying goals, desires, needs and motivations (Pierre, 2019). It is the direction of consciousness, it is what makes humans *human*, the sense of 'self' and ego (Levy, 2016) and the stream of experience within human cognition and behaviour that controls and enables creativity (Klausen, 2010). Design is not accidental, it is intentional (Galle, 1999). Likewise, deviance is not accidental, it is intentional (Baudrillard, 1996; Cropley et al., 2008; Cropley et al., 2010; Hilton, 2010 Lin, Mainemelis & Kark, 2016). The valence (negative or positive), arousal (low or high) and affect (emotional drive) of the intention and associated emotions (Deona & Scherer, 2010; Russell, 2003) effectively also steer the thinker's thoughts and behavior. This directionality moves towards either promoting the wellbeing of others, or in the case of deviant thinkers, a myriad of unethical scenarios that can vary between the ensuring wellbeing of themselves at the detriment of others, or the deliberate harm of others for numerous highly varied reasons (Machiavellianism [dark personality triad – psychopathy, narcissism, Machiavellianism – indifference to morality], self-gratification, economic need, terrorism, religious extremity and war, revenge etc.). These motives, goals and outcomes vary greatly between deviant thinkers, yet a framework or map and strategy for anticipatory and empathic relations within cybernetic security design would prove extremely valuable.

Here, is where the construct and emotional state of *empathy* is interesting. Empathy is the ability for a person to understand and connect with the mental and emotional lives of others (Read, 2019). Empathy as a construct can be seen as existing very closely to that of intersubjectivity, whereby experience and emotions are generated and exchanged in interaction with two or more individuals (Cooper-White, 2014). Callous acts of deviant thinkers in both on and offline behavior, both are devoid of empathy through the disregard of the target's boundaries (personal space, thoughts, feelings, property, wellbeing), yet at the same time, the means by which these thinkers establish their plans, traps and undertake crime heavily involves the ability to predict the target's behaviour based on anticipated emotions, actions and reactions. That is, while a deviant thinker emotionally detaches themselves from their target(s), they cognitively empathise in a way that ensures they will be far enough ahead in terms of predicting behaviour that they may plot courses of action that will operate in their (the deviant thinker's) advantage. For example, a spouse wishing to undertake insurance fraud through murder, will anticipate and design the death of their spouse based on known and established courses of action and reaction – i.e., routines, defensive behaviour, emotional responses to certain circumstances and the following lives of reactive behaviour. Another example may be a deviant thinker entering into an employment contract in an organization. They will know that some recruiters in highly busy positions will seek answers from perhaps less optimal sources (i.e., trade-off information), will repeat this for example on an employment contract perhaps breaking laws associated with trial periods etc., and will encourage this behaviour in terms of suggestions to the recruiter. Once the recruiter has established a contract based on trade-off (easily accessed yet less than legally optimal) information and it is signed, the deviant thinker holds power in their ability to either sue the organization, and/or do what they like for the duration of the contract knowing full well their employment will not be terminated due to the threat of a law suit.

As seen in these examples, there is an interchange between micro and macro composites of data. The macro composites include matters such as laws, policies and the use of Internet-based information (personal and public domain), and the micro – the information that exists in and around the target's body and psyche. Cybernetic security design aims at not just protecting online information, but offline safety and security. Micro interactions in for instance, face to face situations may and are pivotal in either: a) concretising the 'target' goal or relationship in the mind of the deviant thinker; and/or b) be the means to the end of whatever the deviant thinker would like to execute. As, life is not simply about digitally-bound information, rather it is about people-to-people contact and ecosystems.

From this perspective, while common approaches to design thinking have generally focused on the client, user and their interests, potentially without too much explicit consideration or evaluation of ethical stance (ethics)[1] – how the designed solution will exist, whether or not it represents or potentially generates impact that could be considered good (e.g., socially responsible or ecological friendly) or evil (e.g., maximum profit at the cost of the environment and society's wellbeing) – a lot could be gained through applying a double-layered empathic approach to cybernetic security design. That is, through a pre-emptive and anticipatory model (EMPx2), cybernetic security design both ascertains the thoughts, motivations, emotions, actions and reactions of the 'Target' (the traditional 'weakest link' approach) and combines this understanding with plotting out the thoughts, motivations, emotions, and anticipatory actions of the deviant thinker (see Figure 3).

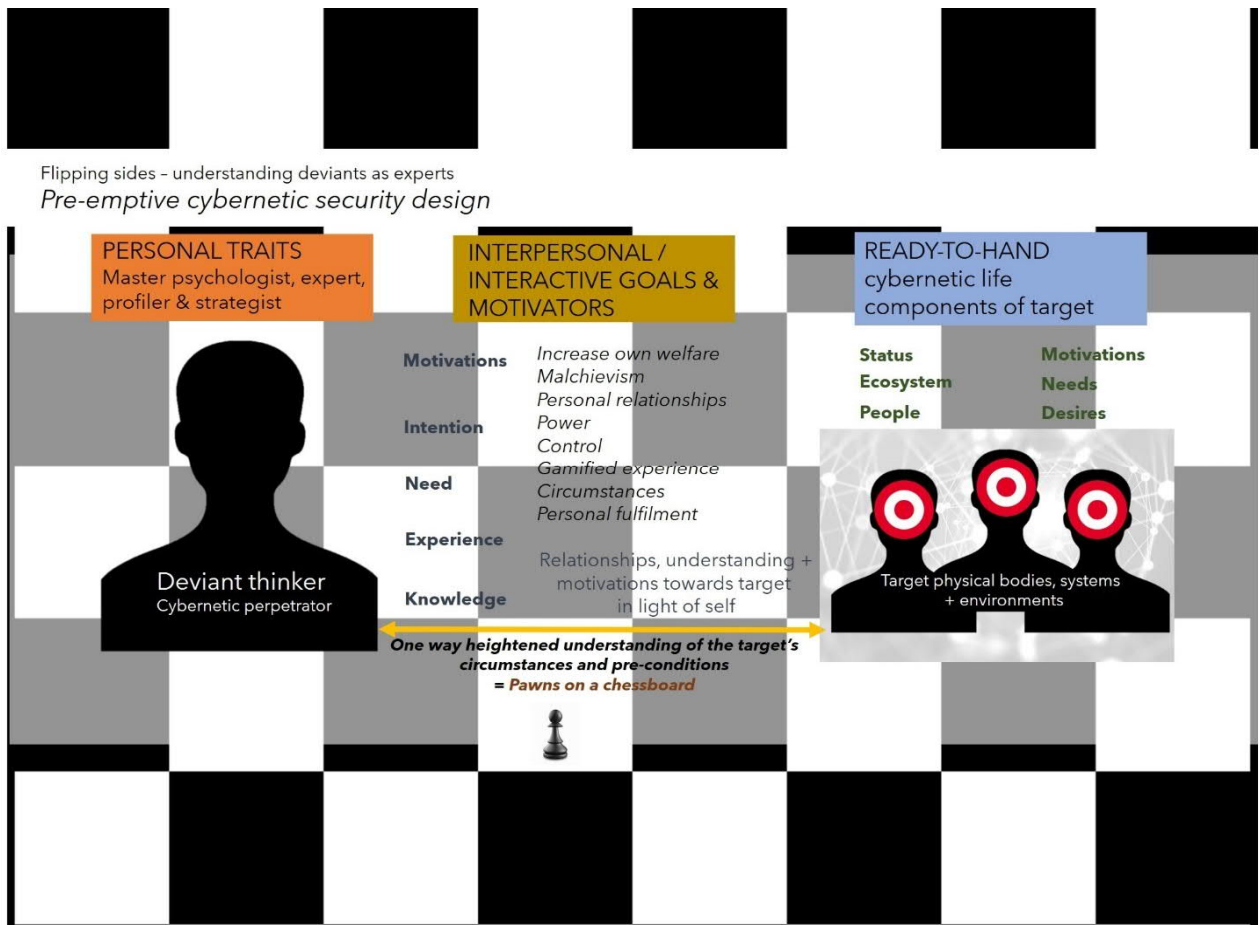

Fig. 3: EMPx2 – double-layered empathic model for cybernetic security design

In EMPx2, deviant thinkers are likened to master chessplayers who analyse the chessboard and its pieces through the lens of motivation, intention, need, experience and knowledge. The deviant with their capacity for creativity, use the pieces (tools, elements and devices available in online-offline environments and e.g., law and policy) that are visible, and some that are not (e.g. psychological resources of the target) to arrange a scenario in which their target is entrapped (check mate). The reason why this cybernetic mode of security design is important is due to the fact that while online data is massive and mostly non-graspable in its expanse, the primary concern of human beings is for their body, its wellbeing and socio-psychological health (Frijda, 2016; Frijda & Mesquita, 1994). For as long as humans have bodies, and for as long as organisations have humans with bodies running them, interpersonal and micro-informational layers are of integral importance. Cyber security covers mainly the digital information side of security, while humans, people, in their face-to-face interpersonal exchanges and dynamics that occur offline remain somewhat exposed in a large hole of vulnerability.

---

[1] Although in this day and age of ethical design, social responsibility and sustainability, increasingly more emphasis is placed on benefits versus costs and moral code within design (see e.g., Baldini et al., 2018; Melles et al., 2011).

In addition to the plurality and diversity among deviant thinkers, there is a complexity of relations between them and their targets, and the expanse of damage that can occur through the victimization of targets. In order to design cybernetically secure solutions with a more effective and anticipatory logic, greater understanding must be attained to isolate the variables within the deviant-target and ecosystemic relationships that render targets susceptible to victimisation. This involves in-depth study of the semiotic systems surrounding and available to deviant thinkers from various streams and in specific circumstances (i.e., insurance fraud, contract manipulation, and highly personal victimization relationships such as those seen in the 'Me too' campaign), and the expanse of creative capacity that any one individual can use to anticipate, manipulate and violate another person's being and property.

The point of this ideas paper was to shed light on the fact that while much effort is being made on designing security for macro (cyber) systems, still human life exists largely in the micro (cybernetic physiological-artificial-social). If truly designing security to protect the lives and interests of humans, then this micro level of target actions is not the only necessary empathic component, but so too is the anticipatory level of deviant thinking, and its mal-empathic calculations. Consideration for the human individual, and smaller scale cybernetic (online-offline) actions and interactions are not simply serving to protect the wellbeing of targets as standalone individuals, but also to protect institutional policies, contracts and wellbeing of full organisations. Furthermore, not only do humans still have bodies, but information technology is quite rapidly merging with human bodily existence, meaning that now perhaps more than ever, designers need to understand the dynamics of the body and intersubjective, micro experiences, to be able to anticipate the factors of security design in future online systems. And, on a final note, it must be remembered that deviant thinkers are not simply faceless scammers sending phishing emails from Internet cafes in developing countries, they are also the friendly (poker) faced people next door. Beware, for although crime is never perfect – Internet or not – "*the corpse of the real has never been found*" (Baudrillard, 1996, p. xii).